\documentclass[10pt,aps,prl,twocolumn,showpacs,superscriptaddress]{revtex4-2}
\usepackage[english]{babel}
\usepackage{amsmath}
\usepackage{amsfonts}
\usepackage{amssymb}
\usepackage{graphicx}
\usepackage{amsthm}
\usepackage{bm}
\usepackage[cp1251]{inputenc}

\begin{document}

\title{Proposal for a Nuclear Light Source.}

\author{E.~V.~Tkalya\thanks{https://orcid.org/0000-0002-3521-969X}}
\email{tkalya\_e@lebedev.ru}
\affiliation{P.~N.~Lebedev Physical Institute of the Russian
Academy of Sciences, 119991, 53 Leninskiy pr., Moscow, Russia}

\author{P.~V.~Borisyuk}
\affiliation{National Research Nuclear University MEPhI, 115409,
Kashirskoe shosse 31, Moscow, Russia}

\author{M.~S.~Domashenko}
\affiliation{P.~N.~Lebedev Physical Institute of the Russian
Academy of Sciences, 119991, 53 Leninskiy pr., Moscow, Russia}
\affiliation{National Research Nuclear University MEPhI, 115409,
Kashirskoe shosse 31, Moscow, Russia}

\author{Yu.~Yu.~Lebedinskii}
\affiliation{National Research Nuclear University MEPhI, 115409,
Kashirskoe shosse 31, Moscow, Russia}

\date{\today}

\begin{abstract}
The paper considers a principal possibility of creating a nuclear light source of the vacuum ultra violet (VUV) range based on the $^{229}$Th nucleus. This nuclear light source can help to solve two main problems --- excitation of the low-lying $^{229m}$Th isomer and precision measurement of the nuclear isomeric transition energy. The Thorium nuclear light source is based on the nuclei implanted in a thin dielectric film with a large bandgap. While passing an electric current through the sample, the $^{229}$Th nuclei are excited to the low energy isomeric state $3/2^+(8.19\pm0.12$~eV) in the process of inelastic scattering of conduction electrons. The subsequent spontaneous decay of $^{229m}$Th is followed by the emission of $\gamma$ quanta in the VUV range. The luminosity of the Thorium nuclear light source is approximately $10^5$~photons/s per 1~A of current and per 1~ng of $^{229}$Th. The suggested scheme to obtain $\gamma$ radiation from the $^{229m}$Th isomer can be considered as a kind of nuclear analogue of the optical radiation from the usual metal-insulator-semiconductor (MIS) junction.
\end{abstract}

\pacs{23.35.+g, 25.30.Dh, 27.90.+b}
\maketitle

The $^{229}$Th nucleus has a unique low-lying isomeric state with the energy $E_{\text{is}}$, which is equal, according to the latest data, to $8.19 \pm 0.12$~eV \cite{Peik-21}. The history of $E_{\text{is}}$ measurements made in the last five decades, from the first values of $E_{\text{is}}\leq 100$~eV \cite{Kroger-76}, $E_{\text{is}} = 1\pm4$~eV \cite{Reich-90} and $3.5\pm1.0$~eV \cite{Helmer-94} via the values $7.6\pm0.5$~eV \cite{Beck-07} and $7.8\pm0.5$~eV \cite{Beck-R} to the currently accepted values $E_{\text{is}} = 8.30\pm 0.92$~eV \cite{Yamaguchi-19}, $E_{\text{is}} = 8.28\pm 0.17$~eV \cite{Seiferle-19}, $E_{\text{is}} = 8.10\pm 0.17$~eV \cite{Sikorsky-20}, and $8.19\pm0.12$~eV \cite{Peik-21}, has already been described repeatedly and in great detail in various works (see, for example,\cite{Yamaguchi-19,Sikorsky-20,Tkalya-20-PRL} and references therein).

Interest in the isomeric level in $^{229}$Th lies primarily with the possibility of creating ultraprecise nuclear clock \cite{Peik-03,Rellergert-10,Campbell-12,Kazakov-12,Peik-15,Tirolf-19,Beeks-21} and a $\gamma$-ray laser of optical range \cite{Tkalya-11} based on the inversion population of the levels in the low-lying doublet in $^{229}$Th \cite{Tkalya-13,Tkalya-22-NPA}. The nuclear ultra-stable optical clock is of interest to a number of fields of physics. With its use we will be able to do research into the effects of the Einstein equivalence principle violating and the local Lorentz invariance violating \cite{Flambaum-16}, to study of the relative effect of the variation of the fine structure constant $e^2$ and the strong interaction parameter $m_q/\Lambda_{QCD}$ \cite{Flambaum-06,Litvinova-09,Fadeev-20}, to search for dark matter \cite{Safronova-18-RMP}. Alongside these important challenges, the low-lying isomer of $^{229}$Th allows us to explore the exotic process of decay and excitation of the nucleus via an electron bridge \cite{Strizhov-91,Tkalya-92-JETPL,Tkalya-92-SJNP,Kalman-94,Tkalya-96,Porsev-10-PRL,Muller-19,Borisyuk-19-PRC,Dzyublik-20}, to accelerate the $\alpha$ decay of the $^{229}$Th nucleus \cite{Dykhne-96}, to control the isomeric level of $\gamma$ decay via the boundary conditions \cite{Tkalya-18-PRL}, to detect $^{229}$Th ground state decay into the isomeric level in the $(\mu^-_{1S_{1/2}}{}^{229}$Th$)^*$ muonic atom \cite{Tkalya-16-PRA}, to investigate the $^{229m}$Th decay through the metal conduction electrons \cite{Tkalya-99-JETPL}, to observe the Zeno effect \cite{Tkalya-22-NPA} and many others.

In this paper, we propose an algorithm for the solution of one main problem in this field of modern science, namely, the excitation of the $^{229m}$Th isomer in the solid state (dielectric matrix) by electron current.

The most efficient method of excitation by resonant laser photons in the process of direct photoexcitation of $^{229m}$Th \cite{Tkalya-96} or excitation via the electron bridge \cite{Tkalya-92-JETPL,Tkalya-92-SJNP,Kalman-94,Tkalya-96,Porsev-10-PRL,Bilous-18-NJP,Muller-19,Borisyuk-19-PRC,Bilous-20} is currently impossible. These methods require precise tuning of laser radiation to an unknown wavelength of the nuclear transition or stimulation of an atomic $M$1 transition close in energy to the nuclear transition in the process of nuclear excitation by electron transition (NEET see in \cite{Morita-73,Tkalya-92,Karpeshin-17}), which is the ``second stage'' of the  excitation of $^{229m}$Th via the electron bridge. However, it should be noted that the excitation of the $^{229m}$Th isomer in the process of direct photoexcitation of the 29~keV level, first proposed and theoretically substantiated in \cite{Tkalya-00-PRC}, has already been successfully implemented experimentally \cite{Masuda-19}.

Registration of photons from the decay of $^{229m}$Th could immediately determine the energy of nuclear transition with high accuracy in a low-lying doublet of levels of the $^{229}$Th nucleus and open the way to resonant pumping of the isomer. First successful observations of the gamma decay of the $^{229m}$Th isomer accompanying the beta decay of $^{229}$Ac, implanted into a large-bandgap dielectric, was reported in Ref.~\cite{Kraemer-22}. Another type of experiment to observe the gamma decay of $^{229m}$Th (produced by the alpha decay of $^{233}$U) in an ion trap is described in \cite{Seiferle-22}.

The present work discusses another way of creating a source of $\gamma$ quanta, which accompanies isomeric decay of $^{229m}$Th.

The low-energy isomeric transition in $^{229m}$Th has a high probability of internal conversion \cite{Strizhov-91,Bilous-17,Tkalya-19-PRC-IC_Rydb,Tkalya-20-PRC}. It was the measurements of the conversion electrons energy in works \cite{Wense-16,Seiferle-19} that gave one of the modern values $E_{\text{is}}$. However, on the other hand, the internal conversion makes registration of photons almost impossible. The solution to the problem was described in \cite{Tkalya-00-JETPL,Tkalya-00-PRC,Dessovic-14}. The papers have shown the decay of  $^{229m}$Th in dielectrics with a large bandgap, $\Delta$, to occur precisely with the photon emission and the internal conversion at $E_{\text{is}}< \Delta$ to be forbidden energetically. In a wide-gap dielectric, Thorium participates in chemical bonds by donating electrons. The described efficient ionization increases the binding energies of the remaining electrons. It proves to be sufficient to stop the decay of $^{229m}$Th via the internal conversion channel.

The last missing link, namely, an efficient way of non-resonant excitation of the $^{229m}$Th isomer, was recently proposed in \cite{Tkalya-20-PRL}. The cross sections for inelastic scattering of slow electrons with energies of about 10~eV turned out to have reached values (2--6)$\times10^{-26}$ cm$^2$ depending on the value of the nuclear matrix element. These are rather large cross sections by nuclear standards. It is essential that they make it possible to excite the $^{229}$Th nucleus even if the exact value of the nuclear transition energy is not known. For comparison, the typical effective cross section of resonant photoexcitation of the isomer by laser radiation with a linewidth of $\Delta_{\text{las}}\simeq 10^{-10}$~eV, calculated from the formula $\sigma_{\text{eff}}= (\lambda_{\text{is}}^2/2\pi) [\Gamma_{\gamma}(M1,\text{gr}\rightarrow\text{is})/\Delta_{\text{las}}$], where $\lambda_{\text{is}}=2\pi/E_{\text{is}}$ and $\Gamma_{\gamma}$ denotes the radiative transition width (see below), is (2--8)$\times10^{-20}$~cm$^2$. However, as we noted above, for resonant photoexcitation it is necessary to know the wavelength $\lambda_{\text{is}}$ with high accuracy.

\begin{figure}
 \includegraphics[angle=0,width=0.98\hsize,keepaspectratio]{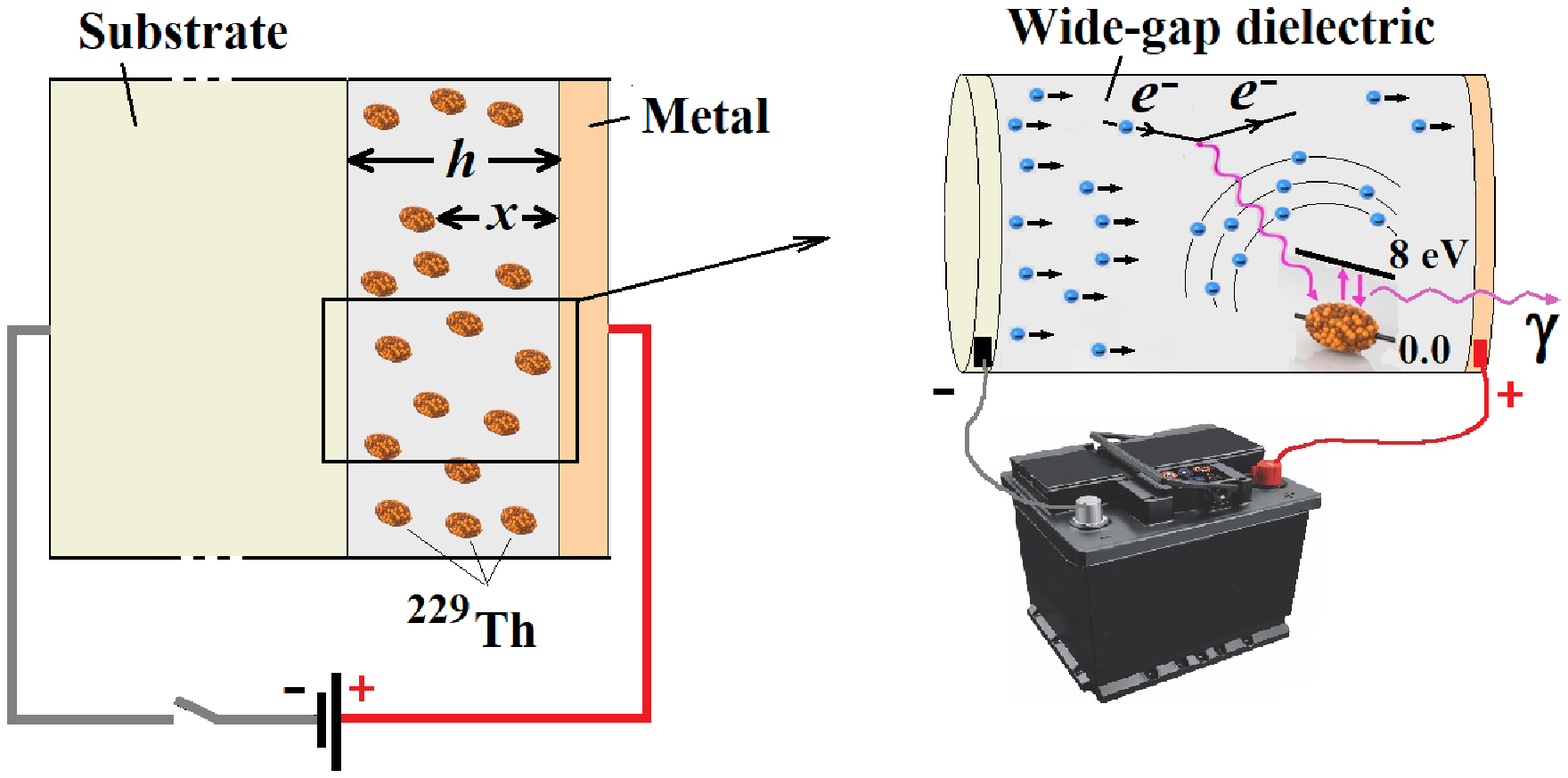}
 \caption{Nuclear light source based on $^{229m}$Th. The left part is a diagram of the device, and the right part is the excitation of $^{229}$Th by the  electrons and the emission of $\gamma$ quantum during the transition of the nucleus to the ground state. ``Metal'' designates a sputtered thin metal contact (Pd and Au were taken for numerical estimates in this work). SiO$_2$ and MgF$_2$ were considered as a large band gap dielectric (they can be replaced by another dielectric with a large band gap). A doped Si and Au were taken as a substrate.}
 \label{fig:LS}
\end{figure}

Now let us see how these problems can be circumvented using a relatively simple scheme. We will consider the layout of the device in Fig.~\ref{fig:LS}. Here $^{229}$Th is implanted into a thin dielectric layer with a thickness $h$ on the order of 10--20~nm with a large bandgap. To date, there are several dielectrics of this sort: SiO$_2$ \cite{Lebedinskii-20}, CaF$_2$, LiCaAlF$_6$,
LiSrAlF$_6$, Na$_2$ThF$_6$, LiYF$_4$ \cite{Rellergert-10}, MgF$_2$ \cite{Pimon-20}. Voltage is applied to the dielectric through a metal contact on one side and a doped semiconductor (Si for axample) or a metal contact on the other side. (If silicon is used as a semiconductor, it is quite easy to grow a thin dielectric SiO$_2$ film directly on a Si substrate. Depending on the doping material, it is possible to prepare an $n$-type or $p$-type semiconductor with the appropriate current flow direction in the device. To be more specific, an $n$-type semiconductor is shown in Fig.~\ref{fig:LS}.)

%
%

As we already noted, the dielectric film should be thin enough to allow passing the electric current. Its optimal thickness is no more than a few electron mean free paths. Thin films of 10-30 nm are convenient for the following reason. The $^{229m}$Th isomer excitation cross section for inelastic electron scattering has a maximum at energies of 9--10~eV \cite{Tkalya-20-PRL}. Therefore, it is reasonable to work with an electrical potential difference between the contacts at a level of several tens of volts (before an electrical breakdown occurs, which will lead to the material degradation. The dielectric strength of SiO$_2$ is approximately 4 GV~m$^{-1}$ \cite{Sire-07}. With the film thickness of 10~nm, the breakdown voltage is 40~V). Under such conditions, the transport of electrons through a thin dielectric film occurs due to the tunneling process (practically, this is the process of electric field electron emission). Thicker films can also be used, but to ensure tunneling, a higher electric potential difference between the contacts should be applied. As a result, the region in which the electron energy is much higher than the optimal one will be significantly increased (see below).

So, the applied voltage is selected so that the electrons scattered by the  $^{229}$Th nuclei have a kinetic energy $E_{\text{kin}}$ which is sufficient to excite the nucleus. In the scheme in Fig.~\ref{fig:LS} $E_{\text{kin}}\sim (q_e/n)V(x/h)$, where $V$ is the applied voltage in volts, dimension of $E_{\text{kin}}$ is electronvolts, $x$ is the depth counted from the electron entry point into the film, $q_e$ is the elementary charge in units of electron charge, $n$ is the refractive index of the dielectric film. Assuming that at the initial time the $^{229}$Th nuclei are distributed over the dielectric layer thickness with some certain density $n_{\text{gr}}(x)$, and that the main voltage drops precisely on the dielectric film, we find the condition necessary for the excitation of nuclei at $x$ depth: $V \geq (n/q_e)(h/x)E_{\text{is}}$. In other words, in order for the excitation reaction of the nucleus to proceed, for example, in half of the dielectric layer, it is necessary to apply the voltage in volts $V = 2nE_{\text{is}}/q_e$, where $q_e=1$.

Let us estimate the $\gamma$ activity of isomeric nuclei excited according to the given scheme. Let the density of the $^{229}$Th nuclei in the target sample at $x$ depth be $n_{\text{gr}}(x,t)$, and the density of the resulting $^{229m}$Th isomers be $n_{\text{is}}(x,t)$. The equations describing the system are as follows:
\begin{eqnarray}
\begin{split}
&dn_{\text{is}}(x,t)/dt=\sigma \varphi n_{\text{gr}}(x,t) -
(\lambda_{\text{tot}}(x)+\tilde{\sigma} \varphi) n_{\text{is}}(x,t),\\
&dn_{\text{gr}}(x,t)/dt=-\sigma \varphi n_{\text{gr}}(x,t) +
(\lambda_{\text{tot}}(x)+\tilde{\sigma} \varphi)
n_{\text{is}}(x,t),
\end{split}
\label{eq:n_is}
\end{eqnarray}
where $\sigma$ is the isomer excitation cross section for inelastic electron scattering in the process of $^{229}$Th$(e,e')$$^{229m}$Th, $\tilde{\sigma}$ is the cross section of the reverse process (calculated from the principle of detailed balance and coincides in order of magnitude with $\sigma$ \cite{Tkalya-20-PRL}), $\varphi=j/S$ is the electron flux density through a target with area $S$ when passing the current $j$, $\lambda_{\text{tot}}(x)$ is the total decay constant or isomeric state width (with using the unit system $\hbar=c=1$) in the sample at $x$ depth. Equations (\ref{eq:n_is(t)}) are solved with initial conditions $n_{\text{is}}(x,0)=0$, $n_{\text{gr}}(x,0)=n_{\text{gr}}(x)$. It should also be noted that the electron flux density of a given energy $\varphi$ is practically independent of $x$ due to the small thickness of the dielectric layer.

Let us simplify the equations Eq.~(\ref{eq:n_is}). 1) We neglect the decrease in the density of the $^{229}$Th nuclei in the ground state with time, that is, let $n_{\text{gr}}(x,t)=n_{\text{gr}}(x)$. 2) We discard the term $-\tilde{\sigma} \varphi n_{\text{is}}(x,t)$, which describes the inverse process  $^{229m}$Th$(e,e')$$^{229}$Th of nuclei deexcitation during inelastic electron scattering. These simplifications are to be made as the condition $\sigma \varphi \tau \ll 1$ is met for the considered current density of the order of 1~A~cm$^{-2}$ (in this case $1/(\sigma \varphi) \gtrsim 10^6$~s) and the expected lifetime of the isomeric state $\tau(x) = 1/\lambda_{\text{tot}}(x) \simeq 10^2$--$10^3$~s (see below). On the one hand, this condition implies that the main decay channel for $^{229m}$Th is the spontaneous decay of the isomeric level, and not the transition to the ground state upon inelastic electron scattering. On the other hand, the density of the excited $^{229m}$Th nuclei at $x$ depth increases according to the characteristic law
\begin{equation}
n_{\text{is}}(x,t)=n_{\text{gr}}(x) \frac{\sigma
\varphi}{\lambda_{\text{tot}}(x)}
\left(1-e^{-\lambda_{\text{tot}}(x)t} \right),
\label{eq:n_is(t)}
\end{equation}
where $\sigma \varphi/ \lambda_{\text{tot}}(x)$ is the relative equilibrium concentration of isomeric nuclei at $x$ depth at times $t\gg \tau$. In order of magnitude $\sigma \varphi/ \lambda_{\text{tot}}(x)\simeq 10^{-4}$, and density $n_{\text{gr}}(x)$ does not change much with time.

Now let us return to the question of the value $\tau(x)$. To do this, we first estimate the characteristic radiative width of the isomeric state $\Gamma_{\gamma}$ in the vacuum
$$
\Gamma_{\gamma}=10\omega_{\gamma}^3\mu_N^2 B_{\text{W.u.}}(M1,
{\text{is}}\rightarrow {\text{gr}})
$$
where $\mu_N$ is the nuclear magneton, $M1$ is the multipolarity of isomeric transition.

In order to calculate $\Gamma_{\gamma}$ it is necessary to know the reduced probability of nuclear transition in Weiskopf units $B_{\text{W.u.}}(M1, {\text{is}}\rightarrow {\text{gr}})$. In work \cite{Tkalya-15-PRC} the value $B_{\text{W.u.}}(M1,
{\text{is}}\rightarrow {\text{gr}}) = 0.03$ was obtained based on the experimental data \cite{Bemis-88,Gulda-02,Barci-03,Ruchowska-06} for the $M1$ transitions between the rotation bands $3/2^+[631]$ and $5/2^+[633]$ in the $^{229}$Th nucleus and the Alaga rules \cite{Dykhne-98_ME,Tkalya-15-PRC}). The other value $B_{\text{W.u.}}(M1, {\text{is}} \rightarrow {\text{gr}}) = 0.006-0.008$ was obtained in \cite{Minkov-17}. It was the result of a computer calculation made in compliance with the modern nuclear models. These two values for $B_{\text{W.u.}}(M1, {\text{is}}\rightarrow {\text{gr}})$ give the time interval $T_{1/2}=\ln(2)/\Gamma_{\gamma}=20$--100 min for the radiative decay of the isomer in Thorium ions in vacuum at the nuclear transition energy $\omega_{\gamma}=E_{\text{is}}=8.2$~eV.

\begin{figure}
\includegraphics[angle=0,width=0.98\hsize,keepaspectratio]{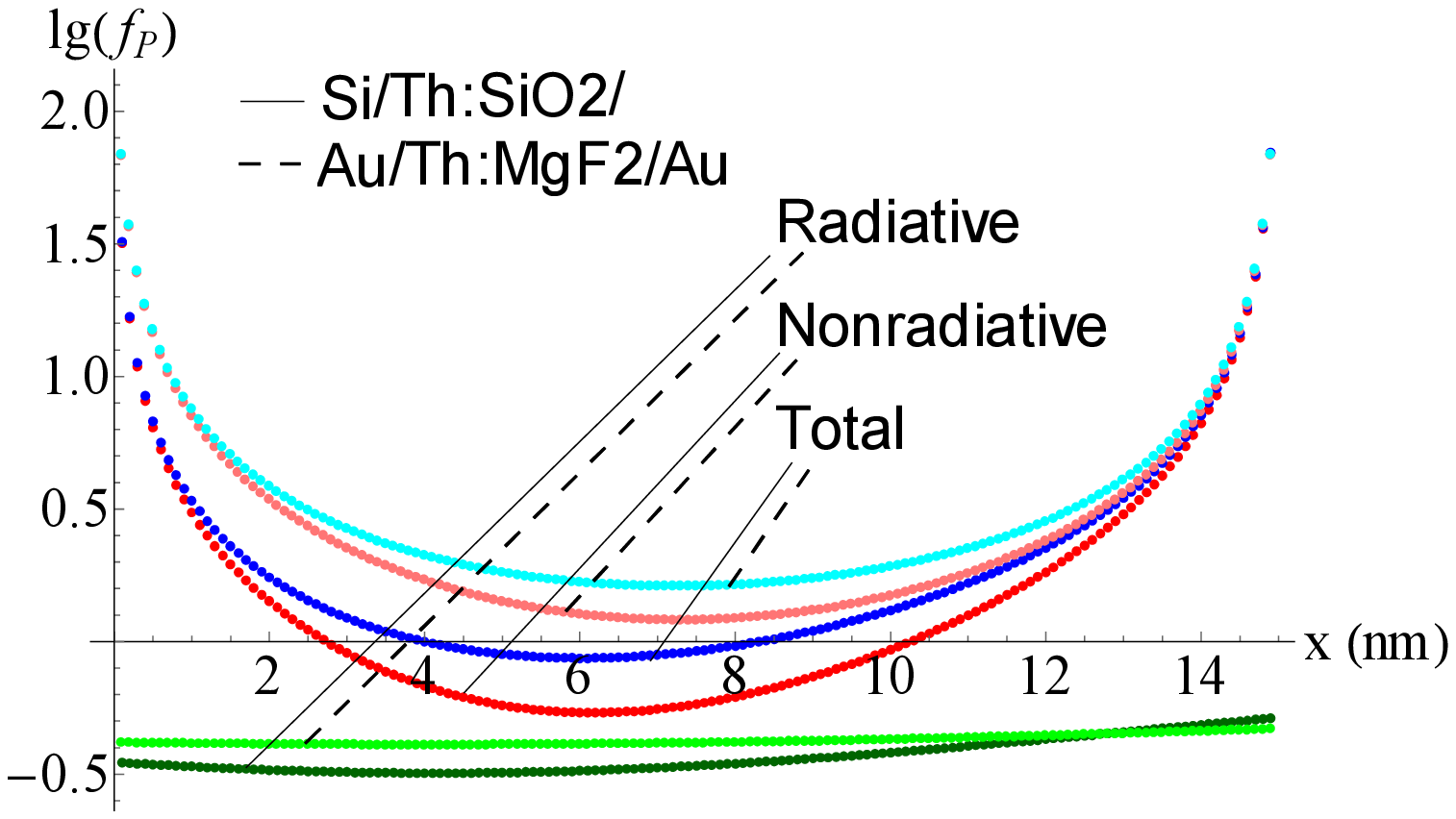}
\caption{The radiative, $f_P^{\text{r}}(x)$, nonradiative,
$f_P^{\text{nr}}(x)$, and total, $f_P^{\text{tot}}(x)=
f_P^{\text{r}}(x)+f_P^{\text{nr}}(x)$ components of the Purcell
factors for the $M1$(8.2~eV) $\gamma$ transition in the scheme in
Fig.~\ref{fig:LS} with $h=15$~nm for thickness of the large band gap dielectric and the metal layer thickness of 10 nm. The thickness of the substrate in calculations is assumed to be infinite.}
\label{fig:Purcell}
\end{figure}

The probability of decay of the $^{229m}$Th isomer in the sample in Fig.~\ref{fig:LS} is influenced by two factors --- the refractive index of the medium \cite{Tkalya-00-JETPL,Tkalya-00-PRC} and the presence of interfaces between the media (the Purcell effect) \cite{Tkalya-18-PRL}. The probability of magnetic dipole ($M1$) $\gamma$ radiation of  $^{229m}$Th in a medium with a refractive index $n$ increases by $n^3$ times compared to the probability of radiation in vacuum \cite{Tkalya-00-JETPL,Tkalya-00-PRC}. The Purcell factor, $f_P(x)$, that is the ratio of the probability of emission in a medium with boundaries to the probability of emission in an infinite medium, affects the decay of $^{229m}$Th more intricately. Fig.~\ref{fig:Purcell} shows the Purcell factors for the $M1$ $\gamma$ transition with the energy of 8.2~eV for the system shown in Fig.~\ref{fig:LS}. The calculation was made within the framework of the approach developed and described in detail in the work \cite{Chance-78}. (We boil down to a graphical representation of the result, since the analytic expressions for $f_P(x)$ for four media are becoming very cumbersome.) Pd and Au were taken as examples of the deposited metal contact. Their characteristics, namely, $n=1.3$ and $\kappa=0.9$ for Pd and $n=1.3$ and $\kappa=1.5$ for Au at the energy of 8.2~eV, are given in \cite{Werner-09}. The thickness of the Pd or Au layers is estimated to be equal to 10~nm. It is limited by the requirement of contact transparency for VUV photons. The fraction of the incident power, propagated through the material with thickness $d$, is given by exp(-$d/l_0$), where $l_0=\lambda_{\text{is}}\kappa/(4\pi)$, $\lambda_{\text{is}} \approx 150$~nm. Using the values of $\kappa$ for Pd and Au from Ref.~\cite{Werner-09} given earlier, we obtain that in a metal layer of thickness 10~nm the VUV photon flux is attenuated by about a factor of 3 for Pd and 2 for Au.

At the SiO$_2$/Metal interface, there are surface states whose levels are located inside the SiO$_2$ band gap. The presence of these levels can lead to an increase in the probability of the isomeric decay of $^{229m}$Th through the internal conversion or electron bridge for the Thorium atoms located within 2--3 monolayers from the interface. Additional losses in the photo output can be avoided by lowering the concentration of Thorium atoms in the near-surface layer so that the doping profile reaches a maximum at a depth of approximately 1 nm (of the order of 3 coordination radii). In the case of a uniform distribution of $^{229}$Th, this problem will affect 10-15\% of the nuclei, reducing the optical output by the same amount.

%
%

Fig.~\ref{fig:Purcell} shows the probability of radiative decay determined by $f_P^{\text{r}}(x)$ to decrease due to the interfaces Pd/SiO$_2$ ($x=0$) --- SiO$_2$/Si ($x=h$) and Au/MgF$_2$ ($x=0$) --- MgF$_2$/Au ($x=h$). However, the total decay probability increases significantly due to the factor $f_P^{\text{nr}}(x)$. It characterizes the nonradiative decay of $^{229m}$Th, which arises from the imaginary part of the dielectric constants of Si, Pd, and Au \cite{Chance-78}. As for the silicon oxide, the imaginary part of the dielectric constants of SiO$_2$ lies in the range of 0.02--0.03 at the energy of 8.2~eV \cite{Zukic-90,Marcos-16}. Thus $\kappa$ is very small for SiO$_2$, and it can be neglected in calculations. For MgF$_2$ $\kappa \approx 3\times10^{-7}$ at 8.2~eV \cite{Zukic-90}.

With respect to these factors it is possible to derive the formula for the total $\gamma$ activity caused by an electric current in the entire sample
\begin{equation*}
Q_{\gamma}(t)=\sigma \varphi S \int_0^h{} dx\, n_{\text{gr}}(x)
\frac{f_P^{\text{r}}(x)}{f_P^{\text{tot}}(x)} \left(1-
e^{-\lambda_{\text{tot}}(x) t}\right) ,
\label{eq:Q_gamma(t)}
\end{equation*}
where $\lambda_{\text{tot}}(x)=n^3 f_P^{\text{tot}}(x)
\Gamma_{\gamma}$.

\begin{figure}
\includegraphics[angle=0,width=0.98\hsize,keepaspectratio]{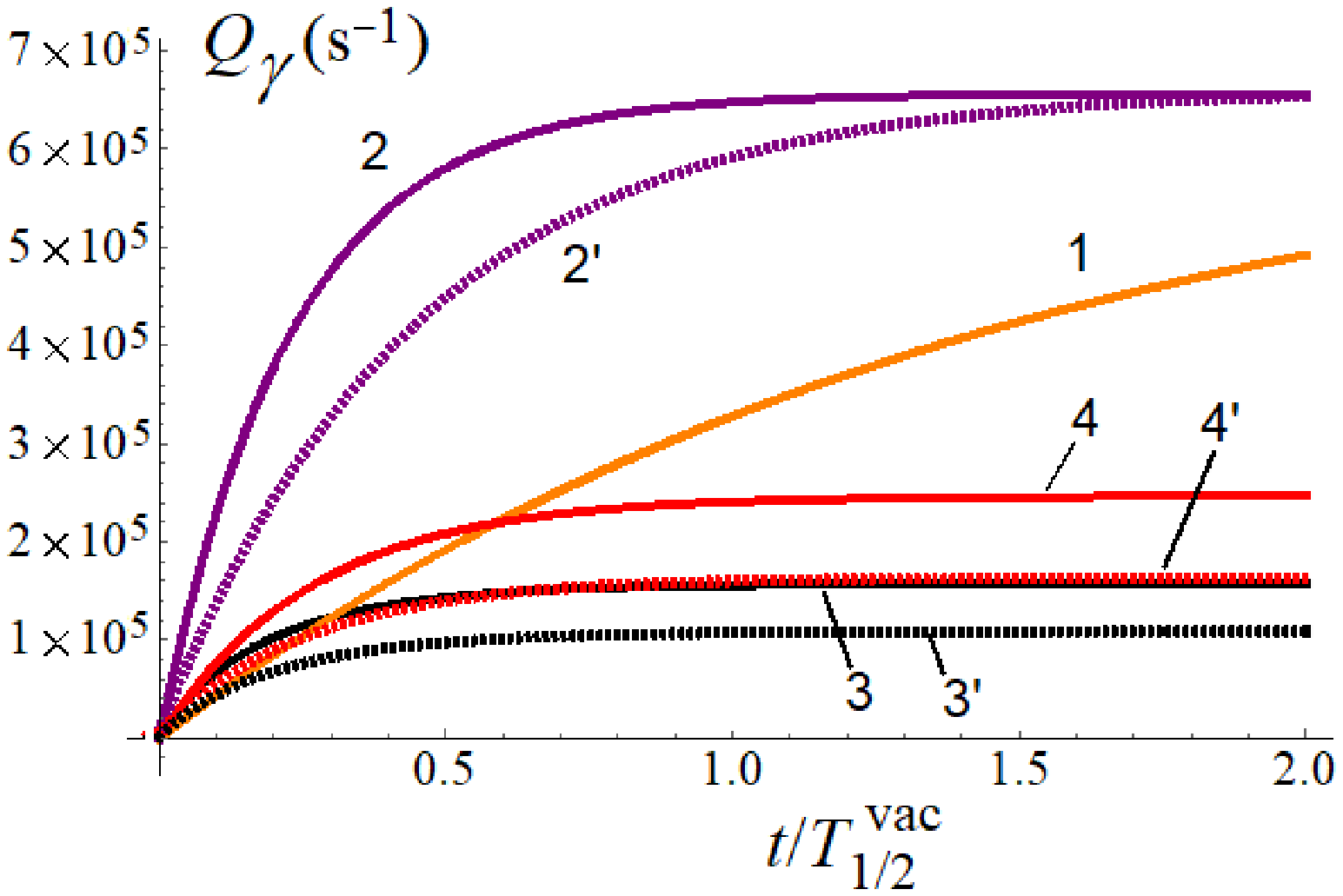}
\caption{The $\gamma$-activity of the sample per 1 A of electric current and per 1 ng of $^{229}$Th as a function of the irradiation time. 1 --- the $^{229}$Th$^{+,++...}$ ion target in the vacuum, 2 (2') --- the same target in the infinite medium of SiO$_2$ (MgF$_2$), 3 (3') --- the $^{229}$Th target in the Si/Th:SiO$_2$/Pd (Au/Th:MgF$_2$/Au) sample in Fig.~\ref{fig:LS} with a uniform distribution of nuclei in depth $n_{\text{gr}}(x)={\text{Const}}=N_{\text{gr}}/(hS)$ ($N_{\text{gr}}$ is the total number of $^{229}$Th nuclei implanted in a dielectric layer), 4 (4') --- all the $^{229}$Th nuclei are located in a thin layer inside SiO$_2$  (MgF$_2$) at a depth of $x\approx 6$~nm, where the functions $f_P^{\text{nr}}(x)$ have a minimum value (this is a kind of upper limit for the $\gamma$-activity for the sample in Fig.~\ref{fig:LS}).}
\label{fig:Qgamma}
\end{figure}

Fig.~\ref{fig:Qgamma} shows the graphs of induced activity. $Q_{\gamma}$ for the sample from Fig.~\ref{fig:LS} is seen to behave much more differently than the $\gamma$ activity of an equivalent target of $^{229}$Th$^{+,++...}$ ions in vacuum or in an infinite medium of SiO$_2$ or MgF$_2$. Both the emissivity and the observed effective half-life of the entire set of isomeric nuclei change significantly.

%
%

As can be seen from Fig.~\ref{fig:Qgamma}, gamma activity in the sample upon reaching equilibrium at times $t\approx{} T_{1/2}/2$ is approximately $10^5$ photons/s per 1~A of electric current and 1~ng of $^{229}$Th.

$^{229}$Th is $\alpha$ emitter with $T_{1/2} = 7880$~yr. Alpha particles particles, decelerated in SiO$_2$ (MgF$_2$), produce a background radiation. Alpha activity of 1~ng of $^{229}$Th is 7~Bk. The mean free path of $\alpha$ particles with the energy 4--5~MeV in SiO$_2$ (MgF$_2$) is about 20~$\mu$m. With a dielectric layer thickness of 15~nm, only $10^{-3}$ of $\alpha$ particles lose energy in the dielectric and give a contribution to the background. (Alpha particles in the Si (Au) substrate and metal contacts do not produce an optical background.) Assuming that the ionization potential is 10~eV, and that all energy of the alpha particles is converted into photons, we get a background of approximately $10^3$ photons per second and per 1~ng of $^{229}$Th purified from daughter nuclides. This means that we are secured by two orders of magnitude for current, until the signal falls to the background level.

Let us compare this estimate with the background from $^{233}$U. To have the same number of the 8.2~eV $\gamma$ quanta as from 1~ng of $^{229}$Th, the $\alpha$ activity of $^{233}$U should be at the level of $10^5\times$50 (taking into account the 2\% isomer population). This requires about 140~mg $^{233}$U. In this case, the host crystal will have a volume of at least $10^{-1}$~cm$^3$. Alpha particles will be completely decelerated in such a crystal and leave in it all their energy. Thus, other things being equal, the background for $^{233}$U exceeds the estimate obtained for 1~ng $^{229}$Th by 8 orders of magnitude. This underlies the advantage of using the proposed scheme with Thorium.

Another source of the background radiation could be the electron flux through the carrier substrate. In this case, the radiation in the vacuum ultraviolet range comes from the processes of exciton recombination. In experiments specifically carried out in 2021, it has been found that in the Si substrate and SiO$_2$ matrix, the emerging background in the wavelength region around 150 nm is very small and only slightly exceeds the intrinsic noise of the detector \cite{Lebedinskii-21}.

It is planned to register the emission line from the isomeric decay of $^{229m}$Th using the same setup that was used to measure the background. The setup device is as follows. A converging short-focus lens made of a material transparent in the VUV range (MgF$_2$) is placed in front of the sample. A focused beam of light falls on the slit of a diffraction spectrometer. The spectrometer is made according to the Paschen-Runge optical scheme, in which the entrance slit of the spectrometer and the spherical diffraction grating are located on the Rowland circle, and the radiation from the entrance slit falls almost normal to the diffraction grating. This scheme has the following advantages: 1 - only one reflection is used, which makes it possible to minimize the loss of incoming radiation, 2 - it is possible to change the grating tilt to capture the desired part of the spectrum, which allows us to measure spectra in a wide range.

The background spectrum and the $^{229m}$Th radiation wavelength are determined from the diffraction pattern using a CCD matrix cooled to a temperature of -20~$^{\circ}$C under vacuum conditions no worse than $10^{-8}$ Torr. With the CCD matrix we can accumulate the signal simultaneously (without scanning) over all channels of the selected region of the spectrum, which in turn allows for measuring the emission spectrum with a high decay rate over time. The maximum luminosity of the spectrometer reaches $10^4$ (i.e. one photon is registered out of every $10^4$ photons arriving at the spectrometer input).

The experiment can be carried out in two modes. 1. Registration and measurement of the spectrum occur directly during the passage of electric current through the sample. 2. Registration occurs after the electron current is turned off and after the completion of the relaxation of electron-hole pairs and other transient processes --- the background sources. In both modes of operation, to take into account the background correctly, it is necessary to carry out measurements on samples without $^{229}$Th. Such a scheme allows us to separate photons originated in the decay of the $^{229m}$Th isomer from the contributions of all other possible electronic transitions in the Th shell. The background accumulation time is required to correspond to the useful signal accumulation time (see in Fig.~\ref{fig:Qgamma}), i.e. to be about $10^3$~s. The described sequential consideration of the signal from samples with $^{229}$Th and $^{232}$Th will single out the desired emission line of the $^{229m}$Th isomer.

Consider the size of the sample in Fig.~\ref{fig:LS} in terms of the allowable concentration of the thorium atoms. Let us take SiO$_2$ as an example, since we have all necessary data for it. According to the results of Ref.~\cite{Borisyuk-18-LPL}, where Thorium ions are implanted into SiO$_2$, Thorium is oxidized to the ThO$_2$ form. The ThO$_2$ band gap is about 6.5 eV. Therefore, the appearance of the ThO$_2$ impurity in SiO$_2$ leads to a decrease in the band gap of the sample as a whole. It was theoretically and experimentally established in \cite{Borisyuk-18-LPL} that, up to a Thorium concentration of 20\%, the band gap in such SiO$_2$+ThO$_2$ dielectric exceeds 8.5 eV. Thus, in a dielectric matrix of the size 1~cm$^2\times$15~nm, about 10~ng of $^{229}$Th can be placed. (Note that in the theoretical work of a group from the Vienna University of Technology, it has also been found that the band gap of a Thorium-doped MgF$_2$ decreases with the concentration of Thorium \cite{Pimon-20}.)

As a result, we obtain the following estimate: at moderate currents of 1~A, the detector quantum efficiency $\gtrsim 10^{-1}$ and the photon collection angle $\approx 2\pi$, the $\gamma$ radiation detector will register about $10^4$--$10^5$~counts/s from such a matrix, which is sufficient for the effect to be reliably detected above the background level.

This scenario is plausible only if there are no additional electron levels in the band gap of the dielectric matrix containing $^{229}$Th. These can be impurity levels, lattice defects, formed in the process of thorium implantation or $\alpha$ decay of the $^{229}$Th nuclei, and others. Such additional electron states of different nature formed inside the band gap in the process of target fabrication can cause a rapid decay of the $^{229m}$Th isomer through an internal conversion channel or via the electron bridge. This can affect negatively the yield of the nuclear light source. The issue of such additional levels was studied in the works \cite{Jackson-09,Rellergert-10,Hehlen-13,Dessovic-14,Jeet-15,Stellmer-15,Borisyuk-18-LPL,Nickerson-20,Kraemer-22}. In particular, as shown theoretically in \cite{Nickerson-20}, a set of electronic defect states in the $^{229}$Th:CaF$_2$ crystal band gap ($^{229}$Th-doped CaF$_2$) leads to excitations via the electronic bridge mechanism, which proceed at a rate 2 orders of magnitude larger than the direct photoexcitation of the isomeric state.

It is important to note that depending on the Thorium concentration dielectrics seem to behave differently. The band gap of a Thorium-doped CaF$_2$ and MgF$_2$ decreases with the concentration of Thorium \cite{Borisyuk-18-LPL,Pimon-20}. On the other hand, some optimism is inspired by the experimental results of Refs.~\cite{Rellergert-10,Jeet-15,Stellmer-15,Borisyuk-18-LPL,Kraemer-22}. They made dielectric matrices LiCaAlF$_6$, CaF$_2$, SiO$_2$, and MgF$_2$ with Thorium without additional states in the band gap. I.e. the presence of a Thorium dopant in the proper concentration in these crystals does not lead to the appearance of additional electronic levels inside the bandgap and the bandgap width $\Delta$ remains quite large exceeding the energy of the isomeric nuclear transition $E_{\text{is}}$. This important issue should always be at the focus of attention in sample making.

Another problem with thin films on substrates is the mismatch between crystal structures and lattice constants of the film and the substrate. For the MgF$_2$/Si system, this effect should be taken into account. However, this is a separate technological problem, which we do not consider here. As for the SiO$_2$/Si system, it is convenient to use the amorphous SiO$_2$ film obtained by the thermal oxidation of silicon. This technology is well established in microelectronics and it can be used to obtain thin silicon oxide films of high dielectric quality with a large band gap ($\simeq 9$~eV)  \cite{DiStefano-71}.

The proposed scheme can be considered as a nuclear counterpart of a conventional light source. Nuclei are excited as a result of inelastic scattering of electrons. This is a non-resonant threshold process, which, most importantly, does not require precise tuning to an unknown nuclear transition energy, but at the same time produces a sufficiently large number of excited nuclei. As a result, there is a real possibility of detecting $\gamma$ quanta from the nuclear isomeric transition in $^{229}$Th.

This research was supported by a grant of the Russian Science Foundation (Project No 19-72-30014).


%

\end{document}